\RequirePackage{ifpdf}
\ifpdf 
\documentclass[pdftex]{sigma}
\else
\documentclass{sigma}
\fi

\numberwithin{equation}{section}

\def\cD{{\cal D}}
\def\cH{{\cal H}}
\def\cI{{\cal I}}
\def\cL{{\cal L}}
\def\cM{{\cal M}}
\def\cN{{\cal N}}
\def\fb{{\mathfrak b}}

\begin{document}

\allowdisplaybreaks

\renewcommand{\thefootnote}{$\star$}

\renewcommand{\PaperNumber}{082}

\FirstPageHeading

\ShortArticleName{Universal Low Temperature Asymptotics
of the Correlation Functions}

\ArticleName{Universal Low Temperature Asymptotics
of\\ the Correlation Functions of the Heisenberg Chain\footnote{This paper is a
contribution to the Proceedings of the International Workshop ``Recent Advances in Quantum Integrable Systems''. The
full collection is available at
\href{http://www.emis.de/journals/SIGMA/RAQIS2010.html}{http://www.emis.de/journals/SIGMA/RAQIS2010.html}}}

\Author{Nicolas CRAMP\'E~$^\dag$, Frank G\"{O}HMANN~$^\ddag$ and
Andreas KL\"UMPER~$^\ddag$}

\AuthorNameForHeading{N.~Cramp\'e, F.~G\"{o}hmann and A.~Kl\"umper}

\Address{$^\dag$~LPTA, UMR 5207 CNRS-UM2,
Place Eug\`ene Bataillon, 34095 Montpellier Cedex 5, France}
\EmailD{\href{mailto:ncrampe@um2.fr}{ncrampe@um2.fr}}

\Address{$^\ddag$~Fachbereich C -- Physik, Bergische Universit\"at Wuppertal,
42097 Wuppertal, Germany}
\EmailD{\href{mailto:goehmann@physik.uni-wuppertal.de}{goehmann@physik.uni-wuppertal.de}, \href{mailto:kluemper@uni-wuppertal.de}{kluemper@uni-wuppertal.de}}

\ArticleDates{Received August 18, 2010, in f\/inal form October 04, 2010;  Published online October 09, 2010}

\Abstract{We calculate the low temperature asymptotics of a function $\gamma$
that generates the temperature dependence of all static correlation
functions of the isotropic Heisenberg chain.}


\Classification{81Q80; 82B23}

\section{Introduction}

Over the past few years the mathematical structure of the static correlation
functions of the $XXZ$ chain was largely resolved. After an appropriate
regularization by a disorder parameter they all factorize into polynomials
in only two functions $\rho$ and $\omega$ \cite{JMS08}. These are the
one-point function and a special neighbor two-point function which, in
turn, can be represented as integrals over solutions of certain linear and
non-linear integral equations \cite{BoGo09}. This resembles much the
situation with free fermions, and what is behind is indeed a remarkable
fermionic structure on the space of quasi-local operators acting on
the spin chain \cite{BJMST08a}. It allows us, for instance, to calculate
short-range correlators with high numerical precision directly in the
thermodynamic limit \cite{BDGKSW08,TGK10a}.

The low temperature asymptotics of $\rho$ and $\omega$ universally
determines the low temperature properties of all static correlation
functions. In this short note we obtain the low temperature asymptotics
in the special case of the isotropic Hamiltonian
\begin{gather}\label{eq:ham}
     \cH=J\sum_j \big(\sigma_{j-1}^x \sigma_{j}^x+ \sigma_{j-1}^y \sigma_{j}^y
        + \sigma_{j-1}^z \sigma_{j}^z\big)
\end{gather}
with no magnetic f\/ield applied and vanishing disorder parameter. Then
$\rho = 1$ and we are left with only one function (and its derivatives)
which, up to a trivial factor, is the function $\gamma$ def\/ined in~\cite{BGKS06}.

\section[Definition of the basic function $\gamma$]{Def\/inition of the basic function $\boldsymbol{\gamma}$}

For our purpose here it is convenient to introduce the function $\gamma$
within the context of a special realization of a six-vertex model
(see e.g.~\cite{BGKS07}) and its associated quantum transfer matrix~\cite{Suzuki85}. By def\/inition the latter has $2(\cN+\cM)$ vertical
lines alternating in direction and carrying spectral parameters
\begin{gather*}
     \underbrace{u,-u,u,-u,\dots,-u}_{2\cN},
     \underbrace{u'+\mu_1,\mu_1-u',u'+\mu_1,\mu_1-u',\dots,\mu_1-u'}_{2\cM}.
\end{gather*}
The spectral parameter on the horizontal line will be denoted $\mu_2$.
We consider this system in the limit $\cN,\cM\rightarrow +\infty$ with the
f\/ine tuning $u\cN=i\frac{J}{T}$ and $u'\cM=i\frac{\delta}{T}$. With an
appropriate overall normalization the largest eigenvalue
$\Lambda(\mu_2,\mu_1)$ is given by
\begin{gather}
     \ln(\Lambda(\mu_2,\mu_1)) = \frac{4\pi J}{T}K(\mu_2)
        +\frac{4\pi \delta}{T}K(\mu_2-\mu_1)\nonumber\\
\phantom{\ln(\Lambda(\mu_2,\mu_1)) =}{}
        +\int_{-\infty}^{\infty}dt
         \frac{\ln\left[(1+\fb(t,\mu_1))(1+\overline{\fb}(t,\mu_1))\right]}
              {2\cosh(\pi(\mu_2-t))}.\label{eq:L}
\end{gather}
Let us note that we recover the familiar system of equations, allowing us
to study the thermodynamical properties of the Hamiltonian~(\ref{eq:ham}),
by setting $\delta=0$. The function~$K(x)$ is def\/ined as
\begin{gather*}
      K(x) = \frac{1}{2\pi} \int_{-\infty}^{\infty}
             dk \; \frac{e^{-ikx}}{1+e^{|k|}}\nonumber\\
\phantom{K(x)}{}
           = \frac{1}{4\pi}\left(\psi\left(1-i\frac{x}{2}\right)
                -\psi\left(\frac{1+ix}{2}\right)
                -\psi\left(\frac{1-ix}{2}\right)
                +\psi\left(1+i\frac{x}{2}\right)\right),
\end{gather*}
where $\psi$ is the digamma function. The auxiliary functions $\fb(x,\mu)$
and $\overline{\fb}(x,\mu)$ are solutions of a~pair of non-linear integral
equations given by
\begin{gather}
     \ln(\fb(x,\mu_1))   =   -\frac{2\pi J}{T \cosh(\pi x)}
        -\frac{2\pi \delta}{T \cosh(\pi (x-\mu_1))}
        +\int_{-\infty}^{\infty}dt K(x-t) \ln(1+\fb(t,\mu_1))\nonumber\\
\phantom{\ln(\fb(x,\mu_1))   =}{}
-\int_{-\infty}^{\infty}dt K(x-t+i) \ln(1+\overline{\fb}(t,\mu_1))\label{eq:b}
\end{gather}
and a similar equation obtained by exchanging~$\fb\leftrightarrow
\overline{\fb}$ and $i\leftrightarrow -i$ in~(\ref{eq:b}). The function
$\gamma$ can now be introduced as
\begin{gather}\label{eq:psi}
     \gamma(\mu_1,\mu_2)=-1+\left(1+(\mu_1-\mu_2)^2\right)T
        \frac{\partial}{\partial \delta}
        \ln(\Lambda(\mu_2,\mu_1))\Big|_{\delta=0}.
\end{gather}

It has been conjectured~\cite{BGKS06} that the correlation functions of the
isotropic Heisenberg chain at any f\/inite temperature (for vanishing magnetic
f\/ield) are polynomials in $\gamma$ and its derivatives evaluated at $(0,0)$.
A similar statement (involving a function $\omega$ and its derivative
with respect to the disorder parameter) was proved for the anisotropic
XXZ chain~\cite{BJMST08a,JMS08,BoGo09}. Amazingly the isotropic limit seems
non-trivial and is still a subject of ongoing work. Here we would only
like to mention that the nearest- and next-to-nearest-neighbor two-point
functions were expressed in terms of $\gamma$ in~\cite{BGKS06} starting
from the multiple integral representation for the density matrix of the
Heisenberg chain obtained in~\cite{GKS05}. The formulae for the longitudinal
two-point functions are, for instance,
\begin{gather}\label{eq:corr2}
     \langle\sigma_1^z\sigma_2^z\rangle_T = -\frac{1}{3}\gamma(0,0),\\
        \label{eq:corr3}
     \langle\sigma_1^z\sigma_3^z\rangle_T =
        -\frac{1}{3}\gamma(0,0)
        -\frac{1}{6}\gamma_{xx}(0,0)+\frac{1}{3}\gamma_{xy}(0,0).
\end{gather}
They will be used below to test our results for the low-temperature
expansion. We denoted derivatives with respect to the f\/irst (resp.\ second)
argument by the subscript $x$ (resp.~$y$). Similar results for four sites
can be obtained from \cite{BDGKSW08} in the isotropic limit. In previous
work~\cite{Tsuboi07} the high-temperature expansion (up to order~25) of
the two-point functions was obtained analytically based on~(\ref{eq:corr2})
and~(\ref{eq:corr3}).

\section{Low-temperature expansion}

To compute the low-temperature expansion of $\gamma$, we follow the line
of reasoning of the article~\cite{Kluemper98}, where a similar task was
performed for the free energy. There are, however, two dif\/ferences between
the usual equations and the ones used in this note: the additional
driving term in~(\ref{eq:b}) proportional to $\delta$ and the shift
$\mu_2$ in the kernel of the integration in~(\ref{eq:L}).

The computation is based on the introduction of a shift
$\cL=\frac{1}{\pi}\ln\left(\pi\frac{J}{T}\right)$ in the auxiliary
functions:
\begin{gather*}
     \fb_\cL(x) = \fb(x+\cL) \qquad \mbox{and} \qquad \widetilde{\fb}_\cL(x)=\fb(-x-\cL).
\end{gather*}
In the low-temperature limit these functions satisfy
\begin{gather}
     \ln(\fb_\cL(x,\mu_1)) \sim -4e^{-\pi x}
        -4 \frac{\delta}{J}~e^{-\pi(x-\mu_1)}+\cD_\cL(x)\nonumber\\
 \phantom{ \ln(\fb_\cL(x,\mu_1)) \sim}{}
 +\int_{-\cL}^{\infty}\! dt\big[ K(x-t) \ln(1+\fb_{\cL}(t,\mu_1))-
        K(x-t+i) \ln(1+\overline{\fb}_\cL(t,\mu_1))\big],\!\!\label{eq:bL}
\end{gather}
where $\cD_\cL(x)$ is the rest of the integral which does not contribute
to the low-temperature limit, when the magnetic f\/ield vanishes (see
\cite{Kluemper98}). A similar relation holds with $\fb\leftrightarrow
\overline{\fb}$ and $i\leftrightarrow -i$ exchanged.

In terms of the shifted functions the largest eigenvalue becomes
\begin{gather*}
     \ln(\Lambda(\mu_2,\mu_1)) \sim \frac{4\pi J}{T}K(\mu_2)
        +\frac{4\pi \delta}{T}K(\mu_2-\mu_1)\nonumber\\
\phantom{ \ln(\Lambda(\mu_2,\mu_1)) \sim}{}
        +\frac{T}{J\pi}\int_{-\cL}^{\infty}dt e^{\pi(\mu_2-t)}
         \ln\big[(1+\fb_\cL(t,\mu_1))(1+\overline{\fb}_\cL(t,\mu_1))\big]
         \nonumber\\
\phantom{ \ln(\Lambda(\mu_2,\mu_1)) \sim}{}
 +\frac{T}{J\pi}\int_{-\cL}^{\infty}dt e^{-\pi(\mu_2+t)}
           \ln\big[(1+\widetilde{\fb}_\cL(t,\mu_1))
                    (1+\widetilde{\overline{\fb}}_\cL(t,\mu_1))\big].
\end{gather*}
To evaluate these integrals we compute
\begin{gather*}
     \cI=\int_{-\cL}^{\infty}dt
         \big[ \ln(1+\fb_\cL(t,\mu_1))\ln(\fb_\cL(t,\mu_1))'
         +\ln(1+\overline{\fb}_\cL(t,\mu_1))
          \ln(\overline{\fb}_\cL(t,\mu_1))'\big]
\end{gather*}
using two dif\/ferent methods. Here the prime stands for the derivative with
respect to $t$. First, we compute it explicitly using the change of
variables $z=\ln(\fb_\cL)$ or $z=\ln(\overline{\fb}_\cL)$, respectively,
which results in
\begin{gather*}
      \cI=2\int_{-\infty}^{0}\ln\big(1+e^z\big)dz=\frac{\pi^2}{6}.
\end{gather*}
Second, we replace $\ln(\fb_\cL(t,\mu_1))$ and
$\ln(\overline{\fb}_\cL(t,\mu_1))$ by their scaling limits~(\ref{eq:bL})
and simplify the resulting expression by taking into account that the
derivative of $K(x)$ is odd and contributions by double integrals cancel
pairwise. This way we obtain
\begin{gather*}
     \cI=4\pi\left(1+\frac{\delta}{J}e^{\pi\mu_1}\right)
     \int_{-\cL}^{\infty}dt e^{-\pi t}
     \ln\big[(1+\fb(t,\mu_1))(1+\overline{\fb}(t,\mu_1))\big].
\end{gather*}
The same type of manipulation can be performed for the functions
$\widetilde b$, and a similar result is obtained with $\mu_1$ replaced
by~$-\mu_1$.

Gathering these f\/indings we obtain the asymptotic form of the largest
eigenvalue,
\begin{gather*}
      \ln(\Lambda(\mu_2,\mu_1))\sim\frac{4\pi J}{T}K(\mu_2)
         +\frac{4\pi \delta}{T}K(\mu_2-\mu_1)
         +\frac{T}{24 J}
          \left(\frac{e^{\pi\mu_2}}{1+\frac{\delta}{J}e^{\pi\mu_1}}
         +\frac{e^{-\pi\mu_2}}{1+\frac{\delta}{J}e^{-\pi\mu_1}} \right).
\end{gather*}
Thus, using (\ref{eq:psi}), the function $\gamma$ behaves asymptotically
for small temperatures as
\begin{gather*}
      \gamma(\mu_1,\mu_2)\sim-1+\left(1+(\mu_1-\mu_2)^2\right)
         \left(4\pi K(\mu_2-\mu_1)
         -\frac{T^2}{12 J^2}\cosh(\pi(\mu_1+\mu_2))\right).
\end{gather*}
This is our main result.

Using (\ref{eq:corr2}) and (\ref{eq:corr3}), we obtain the
low-temperature expansion of the longitudinal correlation functions
\begin{gather*}
      \langle\sigma_1^z\sigma_2^z\rangle_T \sim
         \frac{1}{3}-\frac{4}{3}\ln(2)+\frac{T^2}{J^2} \frac{1}{36},\\
      \langle\sigma_1^z\sigma_3^z\rangle_T \sim
         \frac{1}{3}-\frac{16}{3}\ln(2)+3 \zeta (3)
         -\frac{T^2}{J^2} \frac{1}{36}\left(\frac{\pi^2}{2}-4\right).
\end{gather*}

\begin{figure}[t]
\centering
\includegraphics[width=.60\textwidth,angle=270]{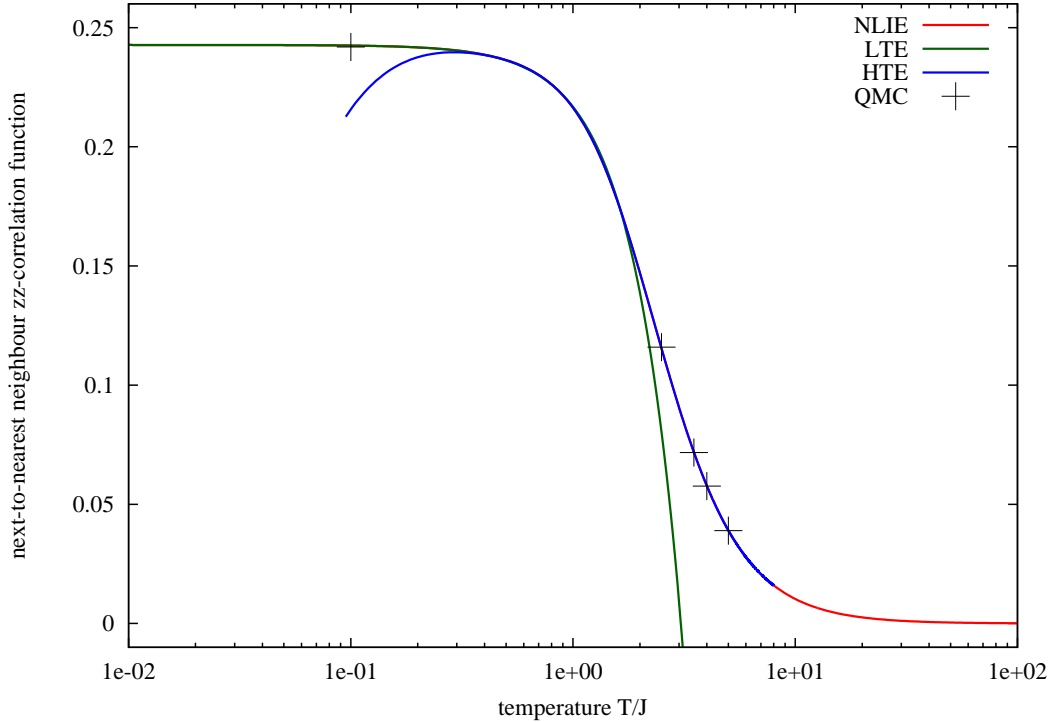}
\caption{Comparison of the high- and low-temperature expansions (HTE, LTE)
of $\langle \sigma_1^z \sigma_3^z \rangle$ with the full numerical solution
obtained from the integral equations (NLIE) and with Monte-Carlo data (QMC).}
\end{figure}

The constant terms (independent of the temperature) in these expansions are
in agreement with those originally found in \cite{Takahashi77,BoKo01}. In
the f\/igure we compare the combined low- and high-temperature
results for the next-to-nearest neighbor $zz$-correlation functions with the
full numerical curve obtained by implementing the linear and non-linear
integral equations that determine~$\gamma$ and its derivatives \cite{BGKS06}
on a computer. The high-temperature data and some additional Monte-Carlo data
are taken from~\cite{TsSh05}. We f\/ind that the numerical curves (NLIE, QMC)
are amazingly well approximated by its low- and high-temperature
approximations.

\subsection*{Acknowledgments}
The authors are grateful to Jens Damerau for providing his computer
program and to Christian Trippe for producing the f\/igure. They wish
to express their gratitude to Z.~Tsuboi and M.~Shiroishi for providing
their data. N.C.\ thanks the department of physics of Wuppertal University,
where this work was initiated, for hospitality.

\pdfbookmark[1]{References}{ref}
\LastPageEnding

\end{document}